\documentclass[runningheads]{llncs}
\usepackage{graphicx}
\usepackage{tikz}
\usepackage{comment}
\usepackage{amsmath,amssymb} % define this before the line numbering.
\usepackage{color}
\usepackage{multirow}
\usepackage[pagebackref=false,breaklinks=true,colorlinks,bookmarks=false,linkcolor=blue]{hyperref}

\usepackage[accsupp]{axessibility}  % Improves PDF readability for those with disabilities.

\begin{document}
% \linenumbers
\pagestyle{headings}
\mainmatter
\def\ECCVSubNumber{7199}  % Insert your submission number here

\title{Super-Resolution by Predicting Offsets:\\ An Ultra-Efficient Super-Resolution Network for Rasterized Images} % Replace with your title

\titlerunning{Super-Resolution by Predicting Offsets}

\author{Jinjin Gu$^{1,\star}$,
Haoming Cai$^2$,
Chenyu Dong$^3$,
Ruofan Zhang$^3$,
Yulun Zhang$^4$,
Wenming Yang$^3$,
Chun Yuan$^{3,5,}$\thanks{Corresponding author.}
}
\authorrunning{Gu et al.}
\institute{
$^1$ The University of Sydney, \quad
$^2$ University of Maryland, College Park, \\
$^3$ Tsinghua Shenzhen International Graduate School, Tsinghua University, \\ 
$^4$ ETH Z\"{u}rich, \quad
$^5$ Peng Cheng National Laboratory,\\
\email{jinjin.gu@sydney.edu.au, hmcai@umd.edu,\\
	\{dcy20, zrf20\}@mails.tsinghua.edu.cn, yulun100@gmail.com,\\
	\{yang.wenming, yuanc\}@sz.tsinghua.edu.cn}}
% \end{comment}
%******************
\maketitle

\begin{abstract}
Rendering high-resolution (HR) graphics brings substantial computational costs. Efficient graphics super-resolution (SR) methods may achieve HR rendering with small computing resources and have attracted extensive research interests in industry and research communities. We present a new method for real-time SR for computer graphics, namely Super-Resolution by Predicting Offsets (SRPO). Our algorithm divides the image into two parts for processing, i.e., sharp edges and flatter areas. For edges, different from the previous SR methods that take the anti-aliased images as inputs, our proposed SRPO takes advantage of the characteristics of rasterized images to conduct SR on the rasterized images. To complement the residual between HR and low-resolution (LR) rasterized images, we train an ultra-efficient network to predict the offset maps to move the appropriate surrounding pixels to the new positions. For flat areas, we found simple interpolation methods can already generate reasonable output. We finally use a guided fusion operation to integrate the sharp edges generated by the network and flat areas by the interpolation method to get the final SR image. The proposed network only contains 8,434 parameters and can be accelerated by network quantization. Extensive experiments show that the proposed SRPO can achieve superior visual effects at a smaller computational cost than the existing state-of-the-art methods.
\end{abstract}

\section{Introduction}
\label{sec:intro}
With the popularity of 4K or even 8K display devices, rendering graphics at ultra-high resolution and high frame rates has attracted extensive research interests in industry and research communities.
However, achieving such a goal is very challenging, as rendering graphics at high resolutions will bring huge computational costs, which poses severe challenges to graphics algorithms and graphic computing devices.
Especially for mobile devices, the huge computational cost also means high power consumption.
Finding ways to enjoy high resolutions and high frame rates without compromising graphics quality as much as possible has become an imperative issue.

In addition to compromising graphics effects and texture quality, there is a new paradigm in recent years to render low-resolution (LR) images first and then super-resolve them to obtain high-resolution (HR) images.
NVIDIA Deep Learning Super Sampling (DLSS) technology and AMD FidelityFX Super Resolution (FSR) are two representative solutions that follow this new paradigm.
However, the poor visual effect and computationally unfriendly features of these super-resolution (SR) methods restrict the application of such techniques.
For example, FSR uses traditional filtering methods to upsample and sharpen LR images.
This method is relatively easy to deploy on various graphics computing devices, but its visual effect is far from ideal.
DLSS uses state-of-the-art deep learning-based SR methods to obtain better perceptual effects, but deep networks' high computational cost makes such a solution only available on dedicated acceleration devices called NVIDIA Tensor Core.
There is no solution available for other devices that do not have such powerful neural computing power.

Deep learning-based SR methods (SR networks) directly map LR rendered images to HR rendered images with many stacked convolutional layers/filters.
One of the reasons for its success is to use a large number of learnable parameters in exchange for sufficient model capacity.
Some successful SR networks even contain tens of millions of parameters, e.g., EDSR \cite{lim2017enhanced}, and RRDBNet \cite{wang2018esrgan}.
The state-of-the-art efficiency-oriented SR network also contains more than 200k parameters, far short of real-time SR requirements.
An efficient deep model can be very ``shallow'' for widely deployed graphic devices, i.e., only three or fewer convolutional layers.
However, within this constraint, deep learning has no precedent for success.

In this paper, we propose an ultra-efficient SR method for computer graphics, especially for mobile platforms and their graphics applications.
Some observations inspire our approach.
Firstly, we argue that generating rich, detailed textures for small SR networks is difficult.
In this case, keeping the edges of the SR image sharp is the most effective way to improve the perceptual effect.
Our method is mainly used to generate such sharp edges.
Secondly, SR is generally performed on anti-aliased images (i.e., FSR) because anti-aliased images typically have gradient distributions and visual effects similar to natural images.
It is very straightforward to borrow an algorithm that works on natural images for anti-aliased images.
However, the edges of the anti-aliased images have more complex gradient distributions, and it is difficult for small SR networks to handle such complex gradient changes.
Rasterized images have simpler color changes, obvious truncation of edges, and similar distributions under various rendering engines than anti-aliased images.
These properties allow us to perform SR on rasterized images using simple operations without wasting the model's capacity to generate sharp edges.

In the light of the above discussion, we propose to perform SR directly on rasterized images.
The difference between HR and LR rasterized images usually occurs in the change of edge pixels.
This change is spatially abrupt (the ``jagged'' edges), and it is difficult to make up for this difference with a simple convolutional network.
Our method predicts offsets that shift surrounding similarly colored pixels to their desired positions to compensate for this sharp difference.
The new SR rasterized image obtained by this method retains the characteristics of the rasterized image well and can produce sharp edges.
We fuse these edges with other areas obtained by the interpolation method to get the final output image.
With subsequent anti-aliasing (AA) algorithm, we can obtain super-resolved rendered graphics at a small cost.
Our final method uses only a three-layer convolutional neural network, which can also be quantized and accelerated and run on various devices with 8-bit integer without loss of final effects.
We present extensive experiments, and the final results show sharp edges and good visual performance.

\begin{figure}[t]
    \centering
    \includegraphics[width=0.9\linewidth]{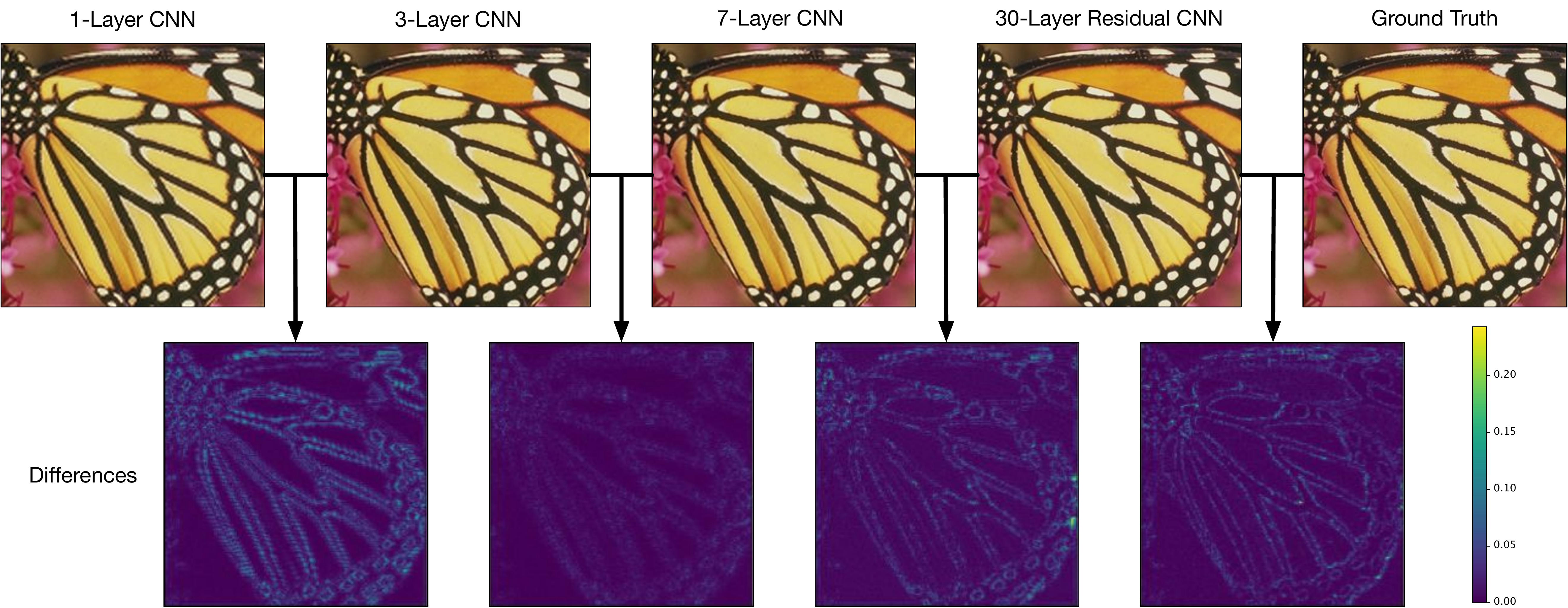}
    \caption{For SR networks, the processing of edges is very important. It can be seen that the 3-layer network mainly improves the edge compared to the network with only one layer. Continuing to increase the network to seven layers did not bring significant changes. Until we deepen the network to 30 layers, the network cannot generate sharper edges. Nevertheless, for the real-time SR problem explored in this paper, how to use a 3-layer network to generate as sharp, visually pleasing edges as possible is the key issue.}
    \label{fig:cnns}
\end{figure}

\begin{figure}[t]
\begin{minipage}[t]{0.48\textwidth}
\centering
    \includegraphics[width=\linewidth]{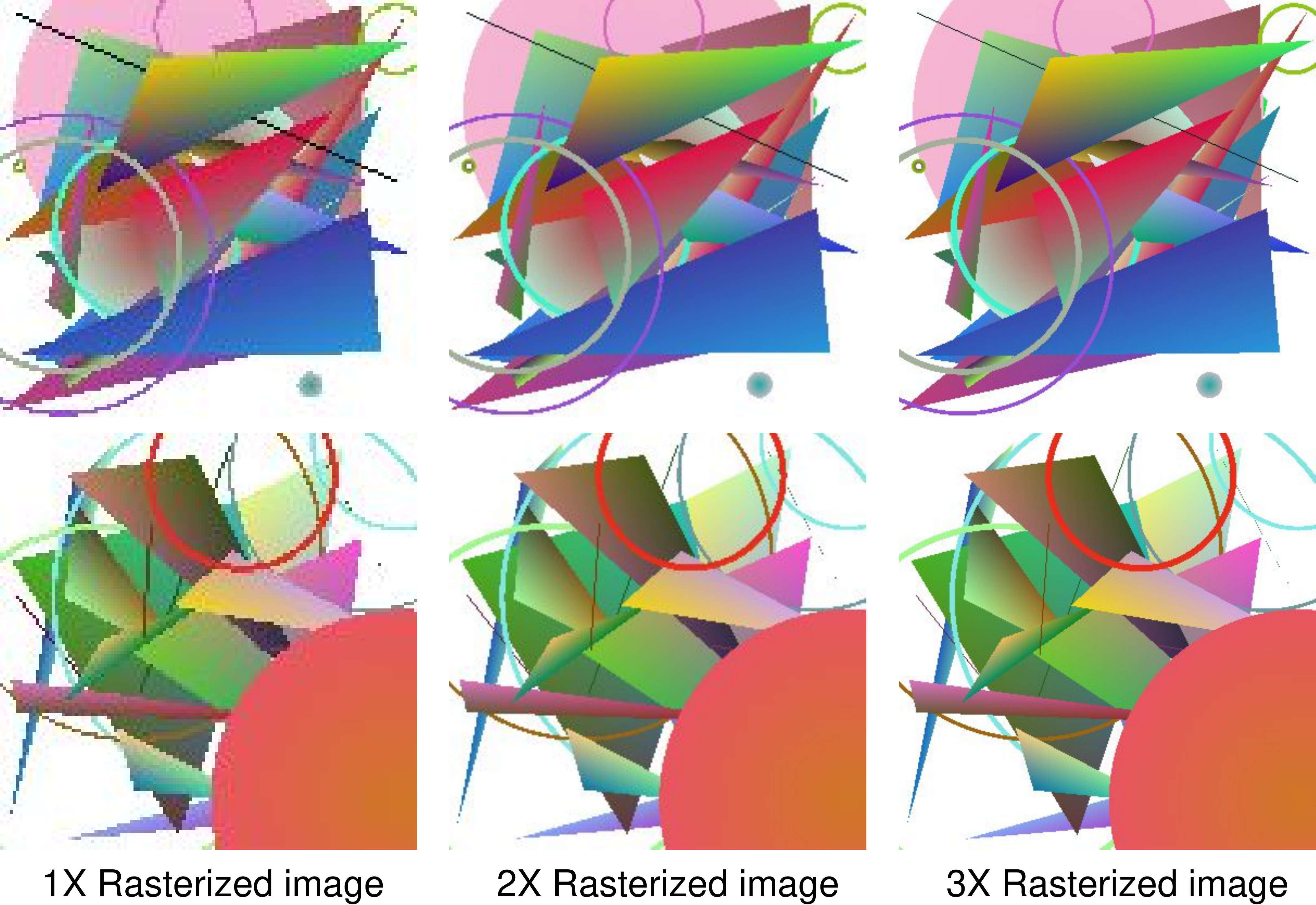}
    \caption{Samples of the generated training images. We rendered 300 images of random geometric objects with random gradient colors, including triangles, circles, rings, lines and bezier curves.}
    \label{fig:data}
\end{minipage}
\hfill
\begin{minipage}[t]{0.48\textwidth}
    \centering
    \includegraphics[width=\linewidth]{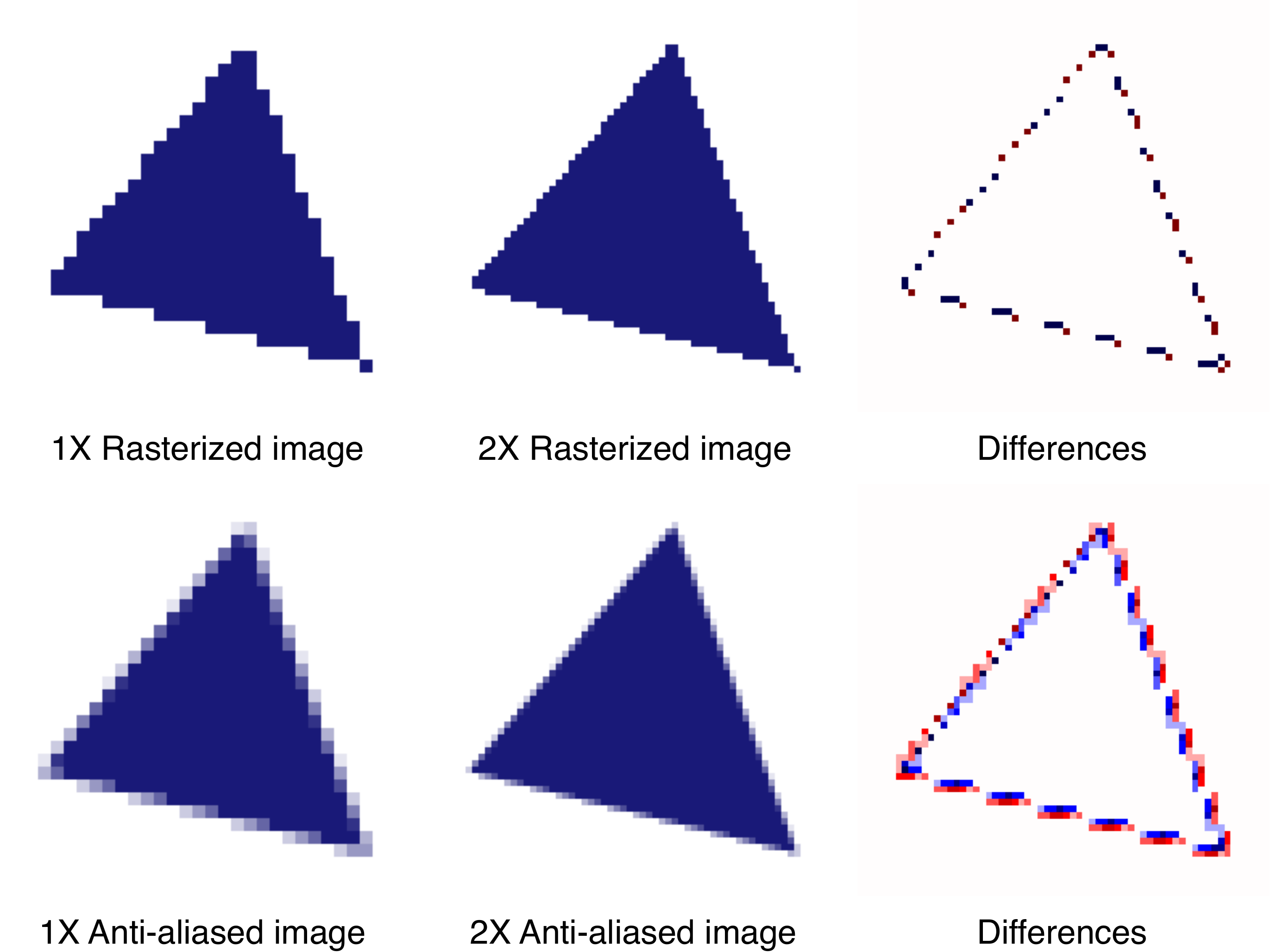}
    \caption{The differences between the rasterized images and the anti-aliased images. The residual images are between the nearest neighbor upsampled LR image and the HR image.}
    \label{fig:diff}
\end{minipage}
\end{figure}

\section{Related Work}
\label{sec:related}

\noindent\textbf{Anti-Aliasing} is a longstanding problem in computer graphics.
Aliasing occurs when sampling is insufficient during rendering because each pixel can only belong to a specific object and receive a unique pixel value, so jagged and saw-toothed effects can appear at the edges of such an object.
Using higher sampling rates is the most straightforward solution to ameliorate aliasing effects, i.e., super-sample anti-aliasing (SSAA) and multisample anti-aliasing (MSAA). 
These approaches are computationally expensive and difficult to be compatible with modern rendering methods such as deferred lighting/shading, and thus not suitable for real-time applications for mobile devices.
Another alternative anti-aliasing paradigm is optimising the visual effects of rasterized images through image post-processing methods.
The most used alternative to MSAA was edge detection and blurring, which is simple, but the results are significantly lower quality than MSAA.
In 2009, Reshetov proposed morphological anti-aliasing (MLAA) \cite{reshetov2009morphological} that performs anti-aliasing with rule-based pixel blending according to certain patterns in the neighborhood.
MLAA has gained extensive attention and applications and inspired many post-processing AA algorithms, e.g., FXAA, SMAA \cite{jimenez2012smaa}, Filmic SMAA and DLAA.
Most of these algorithms have already been applied and created good visual effects.
However, AA algorithms do not increase the resolution of the rendered image, and they can only reduce jagged artifacts at the native resolution.

\noindent\textbf{Super-Resolution (SR)} aims at creating HR images according to the LR observations.
The simplest way to perform SR is through interpolation-based methods that generate HR pixels by averaging neighboring LR pixels, e.g., bicubic, bilinear and Lanczos.
These methods are computationally efficient, but the results are over-smoothed as the interpolated pixels are locally similar to neighbouring pixels and thus have insufficient visual effects.
AMD FSR algorithm first employs a Lanzocs-like upsampling method and uses a sharpening filter to optimize overly smooth edges.
A new fundamental paradigm shift in image processing has resulted from the development of data-driven or learning-based methods in recent years.
Since Dong et. al. \cite{srcnn2014} introduce the first SR network, plenty of deep learning based SR methods have been proposed, including deeper networks \cite{fsrcnn2016,vdsr2016,shi2016real}, recurrent architectures \cite{drcn2016,drrn2017}, residual architectures \cite{lim2017enhanced,srgan2017,wang2018esrgan,li2022blueprint}, and attention networks \cite{rcan2018,san2019,chen2021attention}.
Network-based methods are also used in computer graphics \cite{kaplanyan2019deepfovea,thomas2020reduced}.
More related to this work, Xiao et al. \cite{xiao2020neural} propose a dense video SR network for graphics rendering.
These SR networks can achieve impressive SR results, but the massive parameters and the expensive computational cost limit their practice in real applications \cite{li2022blueprint,kong2022reflash,chen2021attention}.
The development of lightweight SR networks is also a popular research topic.
\cite{shi2016real} propose sub-pixel upsampling operation for SR networks and propose ESPCN as an early attempt in this field.
\cite{ahn2018fast} propose CARN that uses group convolution and implements a cascading mechanism upon a residual network for efficient SR.
IMDN \cite{hui2019lightweight} extracts hierarchical features step-by-step by split operations and then aggregates them by simply using a $1\times1$ convolution
Zhao et al. \cite{zhao2020efficient} construct an effective network with pixel attention, and the entire network contains only 200K parameters.
Despite several years of attempts in this direction, even today's state-of-the-art network is far from being applied to real-time image rendering.
More related to this work, \cite{michelini2022edge} proposes edge–SR (eSR), a set of shallow convolutional layer SR networks that upscale images for edge devices.
This level of computational overhead is likely to be used in real-time rendering.
Thus we will also focus on the comparison with edge-SR in this paper.

\begin{figure}[t]
    \centering
    \includegraphics[width=\linewidth]{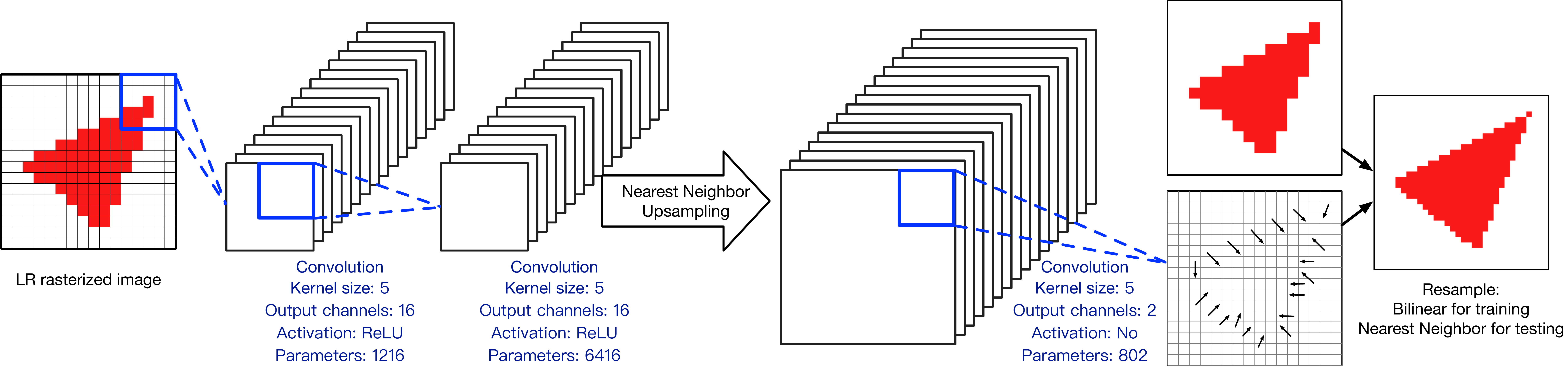}
    % \vspace{-6mm}
    \caption{The proposed SRPO network architecture. The whole network contains only three convolution layers, and no other computationally expensive operation is used. The output of this network is a two-channel offset map instead of a three-channel image.}
    \label{fig:arch}
    % \vspace{-6mm}
\end{figure}

\section{Method}
\label{sec:method}
The methods studied in this paper all have strict computational constraints.
It is difficult for the network to generate so-called ``rich textures'' with a small capacity.
The resulting images are hardly any real improvement over more straightforward interpolation methods \cite{michelini2022edge}.
To generate realistic rich textures, we need to exponentially enlarge the networks we used, or even use generative adversarial models \cite{wang2018esrgan,gu2020image} (see \figurename~\ref{fig:cnns}).
Many works reveal that a human's perception of an image's structure greatly affects the whole image's subjective perception \cite{wang2004image,pipal,gu2020image,gu2022ntire}.
Sharp edges are among the most critical parts affecting this structure's perception.
With limited resources, we argue that enhancing the edges is more appropriate than generating textures in complex and challenging areas, as the interpolation method can already be an economical choice for these areas.
Our method first upsamples the image using interpolation to ensure comparable results in areas other than edges. It then generates images with fine and sharp edges through the proposed network.
At last, we combine the advantages of both through a fusion operation.
We first introduce the proposed SRPO network in Sec~\ref{sec:method:SRPO} and then describe the fusion method in Sec~\ref{sec:method:fusion}.

\subsection{SRPO Network}
\label{sec:method:SRPO}

\subsubsection{Motivation.}
We first describe the motivation for our proposed SRPO network.
The graphics rendering pipeline can be summarized as four stages \cite{akenine2019real}: application stage, geometry processing stage, rasterization stage and pixel processing stage.
The geometric primitive is sampled and discretized into rasterized images in the rasterization stage.
The pixel processing stage involves pixel shading and blending operations.
The post-processing anti-aliasing step is also implemented at the end of this stage.
The current real-time graphics SR methods (e.g., AMD FSR) usually take the anti-aliased images as inputs, as these images have similar edges to natural ones.
Under severe computation constraints, we can only use a minimal number of convolution or filtering operations to process the LR anti-aliased images.
The method at this time does not have a significantly larger model capacity than the linear models, making it very difficult to generate sharp edges.

Our network design begins with the observations of the rasterized and anti-aliased images.
As shown in \figurename~\ref{fig:diff}, the rasterized image shows a jagged-like appearance, while the anti-aliased image has complex grayscale pixels along the edges.
These edges make the image look more natural because they simulate the gradient distribution in natural images.
At higher resolutions, rasterized images still exhibit the same jagged edges, and its anti-aliased images use finer-grained gradients to retouch the edges of the image.
The residuals are equally complex due to the complex edges of both HR and LR anti-aliased images.
This requires SR methods to generate complex residual images to compensate for it.
On the contrary, the residuals of rasterized images are very regular.
We only need to make up for this part of the pixel difference to get an HR rasterised image.
This inspires us to develop an SR method on rasterized images at a small cost.

We further dive into the difference between high and low resolution rasterized images.
When converting an LR rasterized image to an HR rasterized image, we only need to change the image at the edges from background pixel values to foreground pixel values, or vice versa.
The pixel values that need to be changed can usually be found directly from the surrounding -- \emph{we can move the appropriate surrounding pixels to the new positions to complete the upsampling of the rasterized image}.
In this paper, we use a pixel-level offset map to represent the movement of pixels.
Our experiments show that such an offset can be obtained simply by training an ultra-lightweight network.
This leads us to our approach: SR by predicting offsets (SRPO).

\subsubsection{Network architecture.}
We train a simple convolutional neural network $F$ to predict the pixel-level offset map.
This network takes the three-channel LR rasterized image $I^{LR}\in\mathbb{R}^{3\times H\times W}$ as input and outputs a two-channel (x-axis and y-axis) offset map $\Omega=F(I^{LR}),\Omega\in\mathbb{R}^{2\times sH\times sW}$, where $s$ is the SR factor.
The values in $\Omega$ indicate which pixel in $I^{LR}$ the current pixel is composed of.
For example, $(0,0)$ means no offset, that is, the corresponding pixels in $I^{LR}$ are directly used for filling HR results.
$(-1, -1)$ means to use the pixel relative to the lower-left corner of the current pixel.
According to \figurename~\ref{fig:diff}, most of the offsets should be $0$.
The super-resolved rasterized image $I^{SR}_{\Omega}\in\mathbb{R}^{3\times sh\times sw}$ can be obtained by projecting the pixels of $I^{LR}$ according to the offset map $I^{SR}_{\Omega} = P_{\mathrm{nearest}}(I^{LR},\Omega)$.
The subscript $\mathrm{nearest}$ indicates that the projection process uses the nearest neighbor resampling method (the offsets are rounded) to produce similar sharp jagged edges.

The architecture of the proposed network is illustrated in \figurename~\ref{fig:arch}.
The whole network consists of two convolution layers.
The first convolution layer extracts features from $I^{LR}$ and generates 16 feature maps.
The second convolution layer performs a non-linear mapping to these feature maps and generates the other 16 feature maps.
For these two layers, the kernel size is set to $5\times5$, and the activation function is ReLU \cite{glorot2011deep}.
Note that the computation time of such a layer is only about two milliseconds for a 720P input image when using 8-bit integers for low-precision computations on some specially optimized computing devices.
A nearest neighbor upsampling operation is performed on the feature maps produced by the second convolutional layer.
At last, we use another convolution layer to aggregate the upsampled 16 feature maps to the offset map $\Omega$.
The entire network contains 8434 learnable parameters.

\subsubsection{Network Training.}
The proposed network is trained self-super-vised as the ground truth offset maps are unavailable.
The self-supervised learning process is modelled based on the fact that the projected $I^{SR}_{\Omega}$ should be close to the ground truth HR rasterized image $I^{HR}$.
Recall that we want the result of the network could preserve image edges/structure information; the objective function can be formulated using SSIM loss \cite{wang2004image,zhao2016loss}.
The SSIM loss for pixel $[i,j]$ is defined as:
\begin{equation}
    \mathcal{L}(i,j)=\frac{2\mu(I^{SR}_{\Omega})_{[i,j]}\mu_(I^{HR})_{[i,j]}+C_1}{\mu(I^{SR}_{\Omega})_{[i,j]}^2+\mu_(I^{HR})_{[i,j]}^2+C1}\times\frac{2\sigma(I^{SR}_{\Omega},I^{HR})_{[i,j]} + C_2}{\sigma(I^{SR}_{\Omega})_{[i,j]}^2+\sigma(I^{HR})_{[i,j]}^2+C_2},
    \label{eq:loss}
\end{equation}
where the subscript $[i,j]$ indicates pixel index, $C_1$ and $C_2$ are constant, $\mu(\cdot)$, $\sigma(\cdot)$, $\sigma(\cdot,\cdot)$ represent means,
standard deviations and cross-covariance, and we omitted the dependence of means and standard deviations.
The SSIM loss for the whole image is $\mathcal{L}=\frac{1}{s^2WH}\sum_i^W\sum_j^H 1-\mathcal{L}(i,j).$
Means and standard deviations are computed with a Gaussian filter.
The resampling method used in the projection process can not be the nearest neighbor method during training as it produces only discrete results and is non-differentiable.
We use differentiable resampling methods such as bilinear during training to ensure the back-propagation of the network.
Training using Eq~\eqref{eq:loss} can automatically generate reasonable offset maps, although we do not know which offset map is optimal.

For the training data, obtaining real aligned data for training is challenging under most circumstances.
For example, one may need to prepare a lot of aligned HR-LR image pairs for training to enable NVIDIA DLSS in one game.
In this paper, we show that we can use general and simple rendered images for training, and the effect of the model can be well generalized to various game scenarios.
Recall that the network generates sharp edges while upsampling of other areas is done by interpolation methods.
Thus our training data only needs to contain a wide variety of edges.
We rendered 300 images of random geometric objects with random gradient colors, including triangles, circles, rings, lines and bezier curves.
\figurename~\ref{fig:data} shows some image examples used for training.
The LR training images are of size $128\times128$ and the $\times2$ and the $\times3$ HR images are of size $256\times256$ and $384\times384$, respectively.

For optimization, we use Adam optimizer \cite{kingma2014adam} with $\beta_1=0.9$, $\beta_2=0.99$.
The initial learning rate is $3\times10^{-4}$ and halved every 7k iterations.
In the training phase, we set total 40k iterations for our model.
The mini-batch size is set to 16.

\begin{figure}[t]
    \centering
    \includegraphics[width=\linewidth]{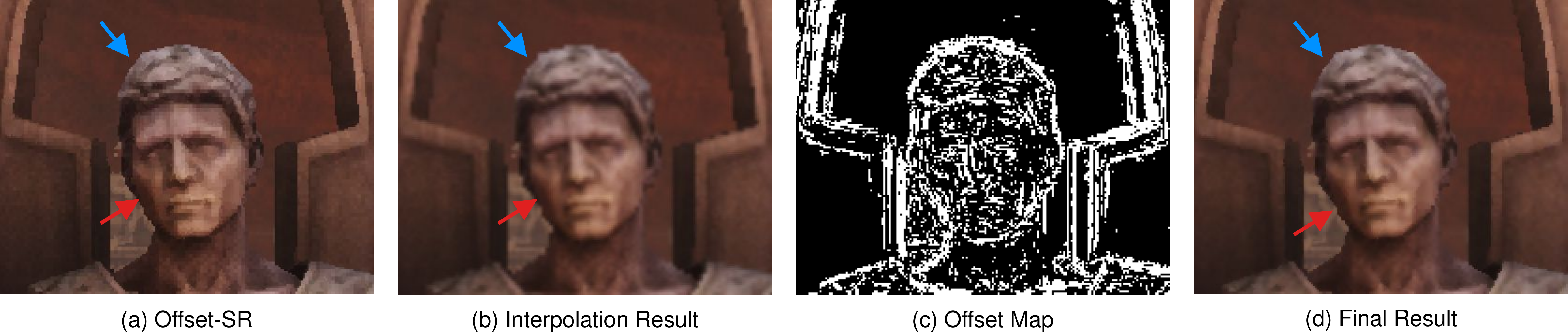}
    % \vspace{-6mm}
    \caption{We combine the edges in the offset-SR $I^{SR}_{\Omega}$ (a) and other areas in the interpolation upsampling result $I^{SR}_{\uparrow}$ (b) through the offset map (c). The final result is shown in (d). As can be seen, there are some non-smooth color transitions in (a) (see the red arrow) because the offset is zero in those places. Its actual effect is the effect of nearest neighbor upsampling. The final effect avoids this while preserving the edges (see the blue arrow).}
    \label{fig:fusion}
    % \vspace{-6mm}
\end{figure}

\subsection{Offset-Guided Image Fusion}
\label{sec:method:fusion}
Although the obtained the SRPO result $I^{SR}_{\Omega}$ can bring images with sharp edges, the rest areas of the image are similar to the effect of nearest neighbour upsampling (the offsets of these areas are close to zero) and cannot be directly used as the final output.
An example is show in \figurename~\ref{fig:fusion} (a).
For convenience, we use the term $I^{SR}_{\Omega}$ as offset-SR.
Simple interpolation methods can obtain reasonable upsampling results for these relatively flat areas.
In this work, we use the Lanczos interpolation method, and we use $I^{SR}_{\uparrow}$ to represent the result.
Next, we compute the linear combination of the corresponding pixels of the two images to fuse the sharp edges of the offset-SR $I^{SR}_{\Omega}$ with other areas of interpolated image $I^{SR}_{\uparrow}$.
The final result can be formulated as $I^{SR}=\alpha\odot I^{SR}_{\Omega}+(1-\alpha)\odot I^{SR}_{\uparrow}$, where the $\alpha\in[0,1]$ is the blending mask, $\odot$ is the element-wise product.
The offset map $\Omega$ contains the movement information for every pixel.
If one pixel is replaced with another value during projection, this pixel belongs to an edge.
We can use the offset map as the blending mask only with some processing:
$
    \alpha_{[i,j]}=\omega_k\otimes\mathbb{I}\big\{\lfloor\Omega_{[i,j]}\rceil\neq0\big\},
$
where $\mathbb{I}\{\cdot\}$ is the indicator function and $\omega$ represent a $3\times3$ Gaussian blur filter with sigma $1.5$.

\begin{table}[t]
    \centering
    \caption{The quantitative results, the amount of floating-point operations and parameters of the different networks involved in our experiments. $\uparrow$ means the higher the better while $\downarrow$ means the lower the better.}
    \label{tab:flops}
    % \vspace{-2mm}
    \begin{tabular}{lccccc}
    \hline
        \multirow{2}{*}{Method} & \multirow{2}{*}{Parameters} & \multirow{2}{*}{FLOPs (G)} & \multicolumn{3}{c}{Metrics}  \\
         & &  & PSNR $\uparrow$ & SSIM $\uparrow$ & LPIPS $\downarrow$ \\
         \hline
        %  EDSR & \color{gray}{1518K} & \color{gray}{1978.2} & \color{gray}{22.67 / 0.7262 / 5.46}  \\
         FSRCNN & 12.64K & 20.54  & 27.26 & 0.8343 & 0.1687\\
         ESPCN  & 8.10K & 6.30  & 27.16 & 0.8195 & 0.1775  \\
         eSR-CNN & 1.97K & 1.53  & 27.28 & 0.8495 & 0.1547 \\
         eSR-TM & 1.57K & 1.22  & 27.20 & 0.8506 & 0.1676 \\
         eSR-TR & 0.6K & 0.47  & 27.22 & 0.8490 & 0.1714  \\
         eSR-Max & 0.036K & 0.03  & 27.03 & 0.8504 & 0.1513  \\
         AMD FSR & - & -  & 27.56 & 0.8639 & 0.1650  \\
         SRPO (ours) & 8.43K & 8.50 & 26.97 & 0.8532 & 0.1501 \\
         \hline
    \end{tabular}
    % \vspace{-6mm}
\end{table}

\begin{figure}[t]
    \centering
    \includegraphics[width=\linewidth]{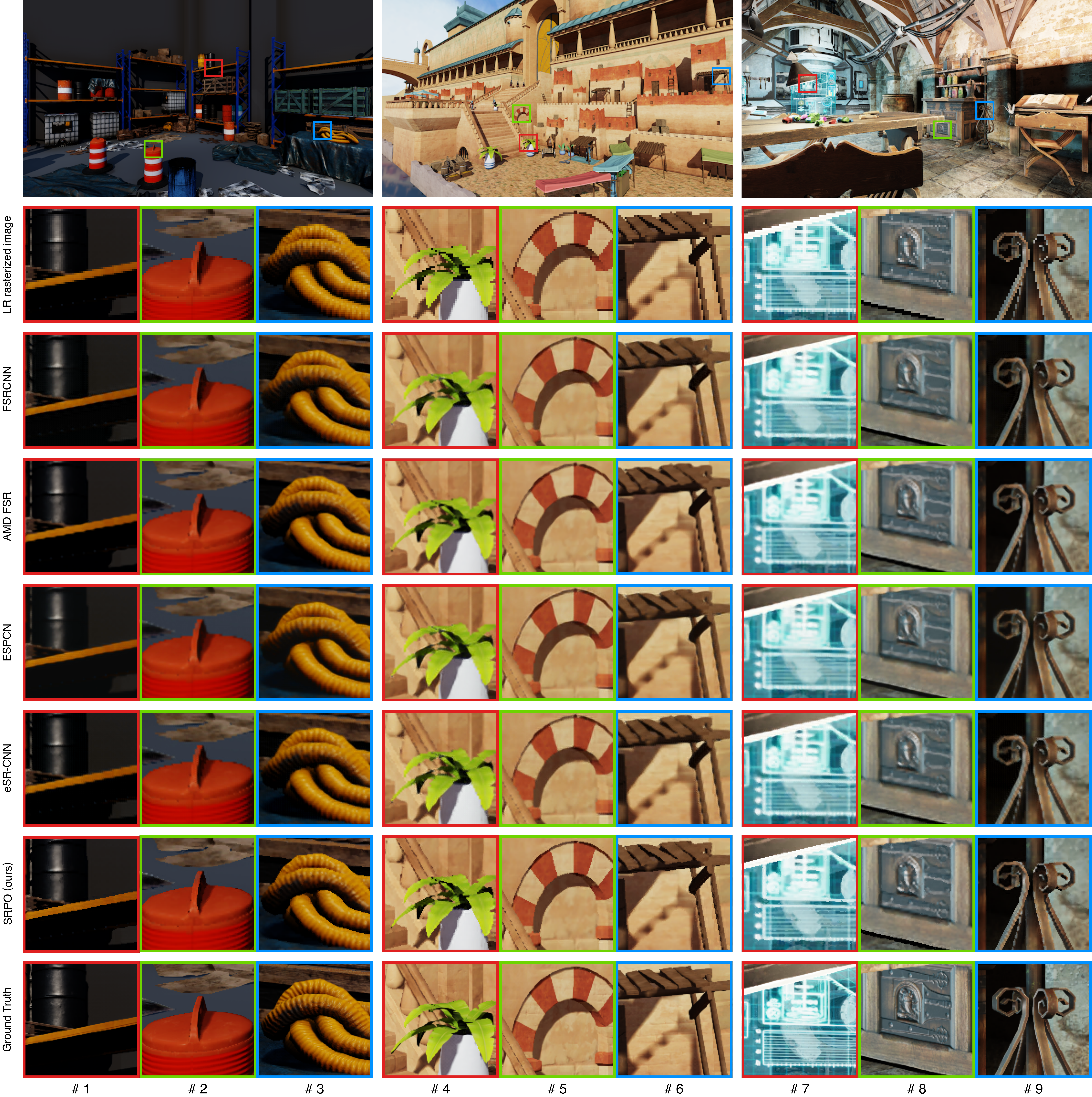}
    \caption{Qualitative comparison of different methods using representative images from our test datasets. The SR factor is 2.
    }
    \label{fig:main_compare}
\end{figure}

\begin{figure}[t]
    \centering
    \includegraphics[width=\linewidth]{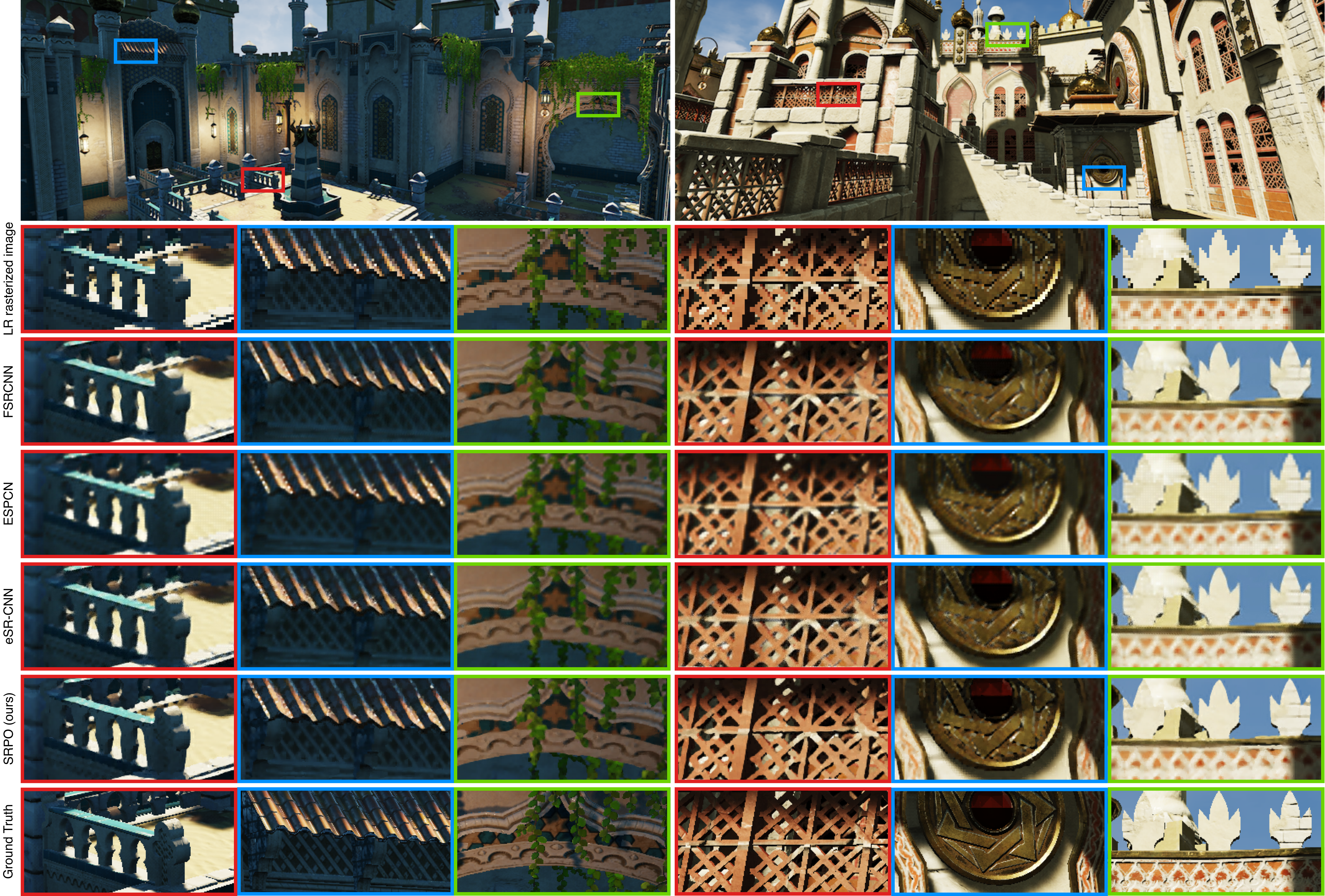}
    \caption{Qualitative comparison of different methods using representative images from our test datasets. The SR factor is 3. Compared with $\times2$ SR, a larger SR factor brings challenges to all the methods. However, the proposed method can still produce sharp edges. This is mainly attributed to its paradigm of predicting offsets.}
    \label{fig:x3}
\end{figure}

\section{Experiments}

\subsection{Comparison}
We prepared our testing data using Unreal Engine 4 with online downloaded scenes and assets.
We render the testing image with one sample per pixel under different resolutions and keep the rasterized and anti-aliased images.
We also applied MLAA anti-aliasing for the HR rasterized images for the final comparison.
All the tests are conducted using tone-mapped images.
There are nine different scenes and hundreds of images, including a wide range of possible scenes in video games, e.g., city streets, gardens, palaces, people, indoor rooms, and spaceship interiors.

We first present a qualitative comparison of the proposed and existing methods.
For the comparison, we include two classic light-weight SR networks FSRCNN\footnote{For FSRCNN, we use the hyper-parameter of $D=56$, $S=12$ and $M=4$} \cite{fsrcnn2016} and ESPCN\footnote{For ESPCN, we use the hyper-parameter of $D=22$ and $S=32$} \cite{shi2016real}, a series of SR networks designed for edge computing called eSR\footnote{For eSR family, we select four versions of eSR: eSR-MAX with $K=3$ and $C=1$; eSR-TR with $K=5$ and $C=2$; eSR-TM with $K=7$ and $C=4$; eSR-CNN with $C=4$, $D=3$ and $S=6$.} \cite{michelini2022edge} and AMD FSR.
Note that FSRCNN includes seven layers and is not designed for real-time processing. We include them only for comparison.
All the trainable methods are re-trained using our paired data for the best performance.
For the AMD FSR method, we use the open-source implementation.
Except for FSR, all the networks are trained and tested using Pytorch framework \cite{paszke2019pytorch}.

\figurename~\ref{fig:main_compare} show some representative regions in our test set.
It can be observed that the proposed method produces sharp edges very close to HR images.
This greatly improves the overall perceptual quality of SR images.
While other methods use neural networks to directly predict output pixels, producing such sharp edges is difficult.
The overall visual effect tends to be blurred.
At some significant jagged edges (see the example \#1, \#2 and \#6 in \figurename~\ref{fig:main_compare} ), the images generated by the other methods still show the jaggedness of the LR image but are only blurred.
The proposed method performs well in these domains to remove aliasing amplification caused by upsampling.
We can also observe that the proposed method still outperforms other methods even on some less apparent edges (without strong color contrast, see \#8 in \figurename~\ref{fig:main_compare}).
The proposed SRPO does not regard these areas as edges that need to recover by moving pixels.
The blending process provides suitable fallbacks for additional robustness.
If the result of SRPO leaves any edge unprocessed (the corresponding offsets are all zeros), The whole method will go back to interpolation upsampling.

Generally, upsampling with a larger SR factor in real-time rendering will significantly drop image quality.
Here we show the comparison results under a higher SR factor of 3 in \figurename~\ref{fig:x3}.
For this experiment, we compare with FSRCNN ($D=56$, $S=12$ and $M=4$), ESPCN ($D=16$ and $S=6$) and eSR-CNN ($C=8$, $D=3$, $S=15$).
The proposed network remains the same architecture as the SR factor 2, only changing the upsampling to $\times3$.
The FSR does not have $\times2$ version. 
As one can see, the proposed method maintains a good performance on the edges and surpasses the existing methods.

We next present the quantitative comparison of different methods.
The comparison results are shown in \tablename~\ref{tab:flops}.
We also show the parameter numbers and floating-point operations \cite{molchanov2016pruning} (FLOPs) of these methods.
Since the optimization of some operations under different hardware environments is different, we use FLOPs to indicate the computational costs for each method without loss of generality.
The FLOPs is calculated through a $3\times1280\times720$ (720p) image.
It can be seen that since we do not optimize using pixel-wise losses such as $l_1$ and $l_2$-norm, the PSNR value of the proposed method is relatively low.
But for the other two indicators, SSIM \cite{wang2004image}, and LPIPS \cite{zhang2018unreasonable}, which are considered to be more relevant to human subjective evaluation \cite{pipal}, the proposed method achieves the level of superior or sub-optimal.
This illustrates the pleasing visual effect of the proposed method.

\begin{figure}[t]
    \centering
    \includegraphics[width=\linewidth]{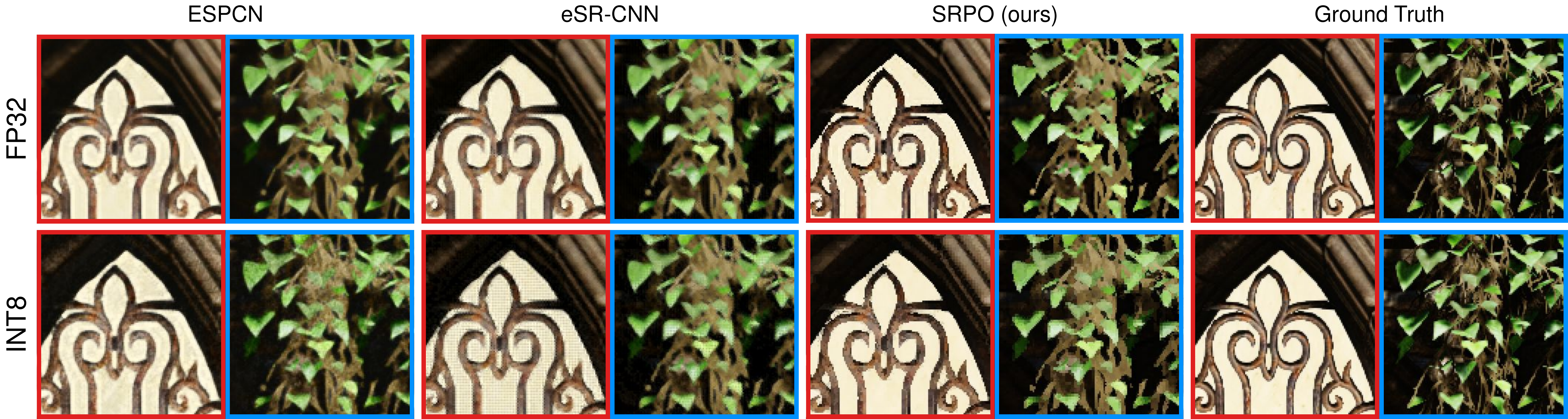}
    \caption{The comparison of quantization effects. For ESPCN and eSR-CNN, the quantization process introduces artifacts that significantly affect image quality. The proposed method is robust to small perturbations because the offset used in the end should be changed to a rounded version.}
    \label{fig:quant}
\end{figure}

\begin{figure}[t]
    \centering
    \includegraphics[width=\linewidth]{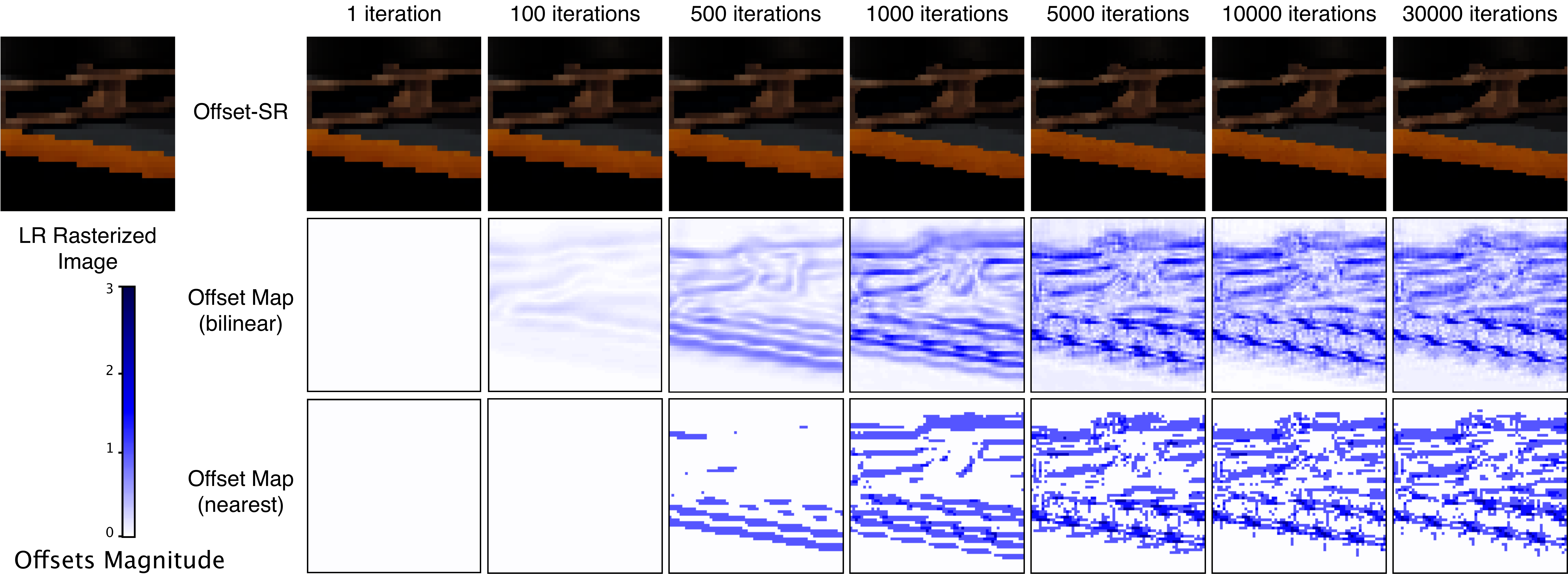}
    \caption{The visualization of the intermediate training results. The offset maps (bilinear) indicate the float-point offsets used during training (bilinear resampling). The offset maps (nearest) indicate the rounded offsets. The color shows the magnitudes of the offset vectors.}
    \label{fig:iters}
\end{figure}

\subsection{Quantization}
The computation of deep learning methods involves intensive operations.
Typically, deep learning training is conducted using 32-bit floating-point numbers.
While efforts are being made to develop efficient SR networks, reducing inference computations to lower bit-depth representations can improve efficiency and reduce power consumption.
For example, half-precision (16-bit) computations tend to be more efficient, and some advanced processing units can support 8-bit integer computations with greater efficiency.
The throughput of 8-bit integer computations is almost twice that of half-precision and 4 times of single-precision.
The good news is that more and more integrated circuits, especially mobile SoCs (system on a chip) have already integrated such advanced processing units, enabling similar technology to be applied on a larger scale.
After training using single-precision (FP32), we apply static quantization \cite{thomas2020reduced} to all the layers of our network.

As shown in \figurename~\ref{fig:quant}, for other methods, quantization brings severe performance degradation and artifacts due to insufficient precision.
There are two reasons for this phenomenon.
First, since the existing network outputs pixel intensity values, the number of intensity values that the network can generate under an 8-bit integer is small.
Artifacts due to loss of accuracy are unavoidable.
Second, since the network involved in this paper is itself a small model, further quantification has a more noticeable impact on the final effect.

The method proposed in this paper is additionally robust to the loss of computing precision because the offset output by our method itself needs to be rounded.
Although the raw output of the SRPO network will also be affected, a certain loss of precision will not significantly change the rounded offset (unless the value itself is around 0.5).
Shifted pixels can retain their original precision (even for 16-bit high-dynamic images).
Therefore, the proposed method is more suitable for quantification.

\subsection{Interpretation of the Learning Process}

We take a deep dive into the learning process of the proposed self-supervised scheme.
Recall that we do not have any prior knowledge about what kind of offset map it should be.
However, the self-supervised training automatically generates the desired offset.
In \figurename~\ref{fig:iters}, we visualize the training of the SRPO network.
We have the following findings.
Firstly, the training of offset is not achieved overnight.
In the initial stage of training, although the output offset-sr does not change, the floating-point part of the offset has been evolving (see 1 and 100 iterations).
The offset that significantly impacts the result during the training process will have a larger loss value, which the network will quickly learn.
For example, the lower boundary of the orange horizontal bar in \figurename~\ref{fig:iters} learns faster than the upper boundary because the gradient of the lower boundary is larger than the upper boundary.
As training progresses, complex structures appear in the optimized offset map.
If the automatic learning of the network does not reveal it, such a complex structure is challenging to be summed up by the rules of manual design.
The self-supervised learning paradigm plays an important role in this work.

\begin{figure}[t]
    \includegraphics[width=\linewidth]{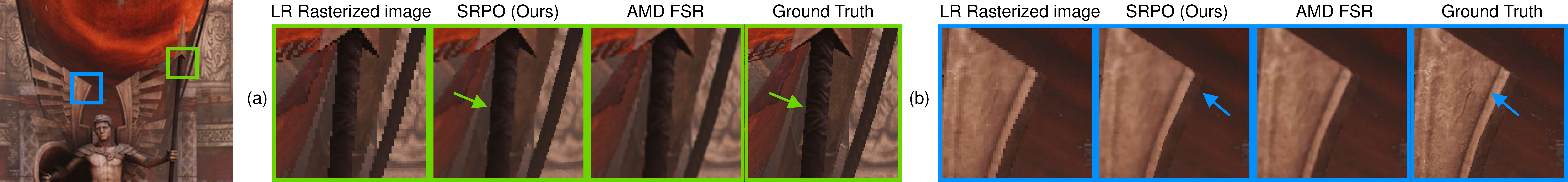}
    \caption{Some failure cases of the proposed method. In case (a), the network cannot identify edges with such slopes due to the limited receptive field. In case (b), the network did not recognize this edge due to the low edge contrast.}
    \label{fig:limitation}
\end{figure}

\subsection{Limitations}
\label{sec:limitation}
The first limitation is the scope of application of the proposed method.
Many advanced graphic technologies in the industry like temporal supersampling, checkerboard rendering, variable-rate shading, or dynamic resolution scaling may affect the ideal condition of the rasterized images assumed in this work.
However, we found that there can be little overlap between these techniques and the goals of this paper.
The techniques described herein are more applicable to mobile devices and their graphics applications.
On these devices, most advanced graphics techniques are difficult to apply.
The proposed method is more suitable for these simpler graphics applications, which are equally important.

Secondly, the limited capacity of the proposed network also brings limitations.
The first problem is that although the network is already very small and can be quantized, it may still not be enough for very high frame rate processing.
We need more advanced network architecture designs, quantization methods, and computing devices to overcome this problem.
Also, due to the limited network depth and receptive field, the results of SRPO still have some aliasing for some difficult edges, see \figurename~\ref{fig:limitation} (a).
This is because detecting a line with such a slope requires a larger receptive field, which is difficult for a network with only three layers to achieve.
Another limitation is the edges with low contrast, as shown in \figurename~\ref{fig:limitation} (b).
In these areas, the network is hard to identify whether the pixel changes are caused by an edge or noise and color gradient.

There are many more challenging sources of aliasing in computer graphics, such as sub-pixel aliasing, thin ropes, thin objects that disappear and appear partially or entirely and shading aliasing.
The proposed method can be expected to suffer from challenges, but these scenarios are also challenging to existing solutions. This paper is still a solid step in this direction.

\section{Conclusion}

In this paper, we propose a new method for real-time rendering.
The proposed method uses an SRPO network to obtain images with sharp edges and then combines these edges with interpolation results.
The SRPO network does not directly generate pixels.
Instead, it moves appropriate surrounding pixels in LR rasterized image to the new positions to form the upsampled rasterized image.
We also investigate the quantization effect of the proposed method.
The proposed method can generate SR results with sharp edges and show good visual quality with less than 10K parameters.

\subsubsection{Acknowledgement.}

This work was partly supported by SZSTC Grant No. JCYJ20190809172201639 and WDZC20200820200655001, Shenzhen Key Laboratory ZDSYS20210623092001004. We sincerely thank Yongfei Pu, Yuanlong Li, Jieming Li and Yuanlin Chen for contributing to this study.

\clearpage

\bibliographystyle{splncs04}
\bibliography{egbib}
\end{document}